\documentclass[apj]{emulateapj}
\usepackage{epsfig}
\usepackage{epstopdf}
\usepackage{graphics}

\begin{document}

\title[]{On the Contribution of "Fresh" Cosmic Rays to the Excesses of Secondary Particles}

\author{Y. Q. Guo, H. B. Hu and Z. Tian}

\affil{Key Laboratory of Particle Astrophysics, Institute of High Energy Physics, Chinese Academy of Sciences,Beijing 100049,China}

\begin{abstract}

  The standard model of cosmic ray propagation has been very successful in explaining all kinds of the 
  Galactic cosmic ray spectra. However, high precision 
  measurement recently revealed the appreciable discrepancy between data and model expectation, 
  from spectrum observations of $\gamma$-rays, $e^+/e^-$ and probably the $B/C$ ratio starting from $\sim$10 GeV energy.
  In this work, we propose that the fresh cosmic rays, which are supplied by the young accelerators 
  and detained by local magnetic field,
  can contribute additional secondary particles interacting with local materials. 
  As this early cosmic ray has a hard 
  spectrum, the model calculation results in a two-component $\gamma$-ray spectrum, 
  which agree very well with the observation. 
  Simultaneously, the expected neutrino number from the galactic plane could 
  contribute $\sim60\%$ of IceCube observation neutrino number below a few hundreds of TeV.  
  The same pp-collision process can account for a significant amount of the positron excesses.
  Under this model, it is expected that the excesses in $\overline p/p$ and $B/C$ ratio 
  will show up when energy is above $\sim$10 GeV.
  We look forward that the model will be tested in the near future by
  new observations from AMS02, IceCube, AS$\gamma$, HAWC and future experiments such 
  as LHASSO, HiSCORE and CTA.

\end{abstract}

\maketitle

\section{Introduction}

  Recent decade has witnessed the great progress made in Cosmic Ray (CR) physics. With new generation 
  of space borne and ground based experiments, CRs are stepping into an era of high precision. 
  PAMELA discovered a clear positron excess for energy between 10-100 GeV in 2009 \citep{2009Natur.458..607A}.
  Recently, the AMS2 collaboration has released its first result, i.e. the positron fraction
  $e^+/(e^-+e^+)$ measurement with energy between $\sim 0.5$ GeV to $\sim 350 $ GeV \citep{2013PhRvL.110n1102A},
  which confirmed PAMELA positron excess with unprecedented high precision.
  These results have 
  stimulated a lot of theoretical studies with point of view from either exotic 
  process \citep{2008PhRvD..78j3520B,2009PhLB..672..141B,2009PhRvD..79b3512Y,2009PhRvD..80b3007Z} or astrophysics 
  process \citep{2009PhRvL.103e1101Y,2009JCAP...01..025H,2009ApJ...700L.170H,2009PhRvL.103e1104B}.

  Due to the fact that anti-proton excess has not been observed, the contribution to the $e^+$ excess from the interaction 
  between CRs and InterStellar Medium (ISM) was generally regarded as unlikely and 
  ignorable \citep{2009PhRvL.102e1101A,2009PhRvL.103h1103B,2009PhRvL.103h1104M}. Later on, 
  high precision observation of diffuse $\gamma$-rays obtained by Fermi-LAT shows that the discrepancy between data 
  and model prediction above $\sim$10 GeV is evident \citep{2012ApJ...750....3A}. 
  Fermi-LAT excess actually is consistent with the 
  multi-TeV excess observed by MILAGRO in inner galactic plane \citep{2008ApJ...688.1078A} by taking into account
  the contribution from fresh CRs\citep{2013ApJ...777..149V}.
    The diffuse $\gamma$-ray in CYGNUS region within tens of degrees observed by 
  Fermi-LAT and MILAGRO \citep{2012A&A...538A..71A} was also explainable by fresh CRs \citep{2009ApJ...695..883B}.
  The continued excess all over the galactic plane and no excess outside the galactic plane does not favour
  the dark matter interpretation. 
  Considering the fast energy loss, the IC scattering process of electron is constrained 
  by earlier study\citep{1995PhRvD..52.3265A}.
  In short, the diffuse $\gamma$-ray excess
  tends to suggest the existence of the extra CRs, which interact with the ISM.
  
  
  The neutrino could only be generated from the interaction between CRs and ISM, 
  which make it a unique probe to study the origin and acceleration of CRs. 
  Thanks to the IceCube experiment, the very high energy neutrino observation has made great
  progress.
  The IceCube collaboration
  reported the detections of two PeV neutrino events and 26 other
  neutrino events from 30 to 400 TeV with 2-year data \citep{2013PhRvL.111b1103A,2013Sci...342E...1I}.
  The number of events exceeds the background
  by 2.8$\sigma$ and 3.3$\sigma$ respectively.
  Recently updated result with a total number of 37 neutrino events from 30 TeV to 2 PeV corresponding to
  5.7$\sigma$ has been published 
  for 3-year data combined \citep{2014PhRvL.113j1101A}.
  By including the TeV energy neutrinos, the spectrum can be described by a power law with
  an index of -2.46.
  Two types of origins have been discussed in literatures: galactic and extra-galactic sources.
   As the extra-galactic contribution might be low to explain 
    the IceCube observation \citep{2010PhRvL.104j1101A,2014arXiv1410.3696T,2014arXiv1412.1690N}. 
    A lot of works focus on the galactic origins, which include the TeV
  $\gamma$-ray point sources \citep{2013ApJ...774...74F,2014PhRvD..89j3002N}, 
  the Galactic center, Fermi bubble region \citep{2013PhRvD..88h1302R,
      2013arXiv1309.4077A,2010ApJ...724.1044S} and the diffuse CR interaction with the 
   ISM \citep{2013APh....48...75G,2014MNRAS.439.3414J,2014ApJ...795..100G}.
    Most recently , Neronov A. and Semikoz D. find that both diffuse $\gamma$-ray 
    and IceCube neutrino excesses can be well described if 
    the spectrum index of galactic CRs is -2.5 \citep{2014arXiv1412.1690N}. As the fresh CRs bear a harder
    spectrum than the one required in \citep{2014arXiv1412.1690N}, its contribution to
     the IceCube neutrino excesses may not be ignorable.

  With all of these high precision results, one can estimate the energy power of the excess particles.
  For simplicity, we assume that the diffuse $\gamma$-rays are all emitted in a distance of 8 kpc from solar system.
  According to the integrated flux of diffuse $\gamma$-ray above 10 GeV, which 
  is 2.5$\times10^{-9}$ $ergs$ $cm^{-2}$ $s^{-1}$ $sr^{-1}$
  \citep{2014arXiv1412.1690N}, the power of excess 
  $\gamma$-ray is $\sim 2\times 10^{37}$ $ergs/s^{-1}$. Similarly, the power of neutrino excess is estimated to be
  $\sim10^{37}$ $ergs$ by adopting IceCube spectra mentioned above and extrapolate the 
  lower energy side to 10 GeV \citep{2014arXiv1410.1749A,2014arXiv1412.1690N}.
  The power of positron excess need to be estimated in a different way. Based on the excess positron flux measured by
  AMS02, its local energy density is $\sim 10^{-5}eV$. Assuming this is the avarage value in 
  the galaxy and considering the volume of our galaxy is $\pi(20kpc)^2(0.2kpc)\sim 10^{67}cm^3$ , 
   the power of excess positron is estimated to be $\sim 4\times 10^{36}$ $ergs/s$ 
  if the positron has a lifetime of $10^6$ years. It is interesting to notice that the energy power of three excess particles is
  in the same order of magnitude
  and thus natural to ask whether these excesses of $\gamma$-rays, 
  neutrinos and positrons share the same origin. If the answer is true, one possible explanation is that one of 
  the hadronic interaction process is missed in the standard model of CR propagation. 
  Following this picture and considering that all excess phenomena happens between 
  $\sim$10 GeV to sub-PeV, the CRs involved
  must have a hard spectrum between $\sim$10 GeV to $\sim$PeV. 
  In addition, the CRs should be efficient in interacting with interstellar 
  medium. Such hard component of CRs and environment do exist in the galactic disk around the young accelerators.
  Traditionally, only the steady state CRs         
  are considered when calculating the diffuse $\gamma$-ray, neutrino and $e^+$. 
  However the steady state solution can not describe the CRs under acceleration and the "fresh" CRs 
  detained by magnetic field around the sources.
  It is understandable that the magnetic field near the source is stronger than the average one, which makes it difficult for the
  CRs to escape. 
  This paper tends to study the contribution of the fresh CRs to secondary particle production.

  The paper is organized in the following way. Section 2 describes the conventional propagation model of CRs and the
  additional secondary production from the fresh CRs interaction with ISM. 
  Section 3 presents the results of the calculation compared with the observation.
  And finally, Section 4 gives the discussion and conclusion.

\section{CR Propagation and Interactions in the Galaxy}

  The expanding diffusive shock, generated in the active phase of astrophisical object 
  such as SNRs \citep{1978MNRAS.182..147B,1978MNRAS.182..443B,1978ApJ...221L..29B}, 
  Galactic Center \citep{1981AZh....58..959P,1981ICRC....2..344S,1983JPhG....9.1139G,2013NJPh...15a3053G}, can accelerate the
  CRs to very high energy. Then
  these particles would diffuse away from the acceleration site, and travel in the galaxy for $\sim 10^7$ years.
  During the journey, the impacts due to the fragmentation and raioactive decay in the ISM results in the production of secondary particles.
  In the meanwhile, the electron suffers energy loss in the interstellar radiation field (ISRF) and magnetic field, which
  may lead to the diffuse $\gamma$-ray emission.
  Those secondary particles would be a good tracer to understand the CR origin and propagation. 
  Considering those effects, the propagation equation can be written as:
\begin{equation}
\begin{array}{lcll}
  \frac{\partial \psi(\vec{r},p,t)}{\partial t} &=& q(\vec{r}, p,t) + \vec{\nabla} \cdot 
                                     \left( D_{xx}\vec{\nabla}\psi - \vec{V_{c}}\psi \right)\\
                                   &+& \frac{\partial}{\partial p}p^2D_{pp}\frac{\partial}{\partial p}\frac{1}{p^2}\psi
                                   - \frac{\partial}{\partial p}\left[ \dot{p}\psi - \frac{p}{3}
                                     \left( \vec{\nabla}\cdot \vec{V_c}\psi \right) \right]\\
                                   &-& \frac{\psi}{\tau_f} - \frac{\psi}{\tau_r}
\end{array}
\label{CRsPropagation}
\end{equation}
  where $\psi(\vec{r},p,t)$ is the density of CR particles per unit momentum
  $p$ at position $\vec{r}$,
  $q(\vec{r}, p, r)$ is the source distribution, 
  $D_{xx}$ is the spatial diffusion coefficient, $\vec{V_c}$ is the convection velocity,
  $D_{pp}$ is the diffusion coefficient in momentum space and used to
  describe the reacceleration process, $\dot p\equiv dp/dt$ is momentum 
  loss rate, $\tau_f$ and $\tau _r$ are the characteristic time scales for fragmentation and
  radioactive decay respectively. The spatial diffusion coefficient is
  assumed to be space-independent and has a power law form 
  $D_{xx}$ = $\beta D_0(\rho/\rho_0)^\delta$ of the rigidity $\rho$,
  where $\delta$ reflects the property of the ISM turbulence. 
  The reacceleration can be described by the diffusion in momentum space 
  and the momentum diffusion coefficient $D_{pp}$ is coupled with the spatial
  diffusion coefficient $D_{xx}$ as \citep{1994ApJ...431..705S}
\begin{equation}
   D_{pp}D_{xx} = \frac{4p^2v_A^2}{3\delta(4-\delta ^2)(4-\delta)w} 
\end{equation}
  here $v_A$ is the Alfven speed, $w$ is the ratio of magnetohydrodynamic
  wave energy density to the magnetic field energy density, which can be
  fixed to 1. The CRs propagate in an extended halo with a characteristic
  height $z_h$, beyond which free escape of CRs is assumed. The values of the
  key parameters of CR propagation are listed in Table \ref{CRPropTable}, which
  is similar with previous studies except the tuned injection spectrum 
  \citep{2009PhRvD..79b3512Y,2010ApJ...720....9Z,2015APh....60....1Y}.

\begin{table}[h]
\begin{center}
\caption{the Propagation Parameters}
\begin{tabular}{cccc}\hline \hline
 $D^0 (10^{28} ~cm^2 ~s^{-1})$   & 5.5  & $\rho_0 ~(GV)$                 & 4  \\
 $\delta$                     & 0.45 & $v_A ~(km ~s^{-1})$             & 32\\ 
 $R_{max} ~(kpc)$               & 20   & $z_h ~(kpc)$                   & 4 \\ \hline \hline
\end{tabular}
\label{CRPropTable}
\end{center}
\end{table}

  
  It is generally believed that SNRs
  are the sources of Galactic CRs. The spatial distribution of SNRs is usually described by following
  empirical formula:
\begin{equation}
  f(r,z) = \left( \frac{r}{r_\odot} \right)^a \exp\left( -b\cdot\frac{r-
  r_\odot}{r_\odot} \right)\exp\left( -\frac{\left|z\right|}{z_s} \right),
  \label{SNRs_Distribution}
\end{equation}
  where $r_\odot=8.5$ kpc is the distance from the Sun to the Galactic
  center, $z_s\approx0.2$ kpc is the characteristic height of Galactic disk,
  $a$=1.25 and $b$=3.56 are adopted from
  \citep{2011ApJ...729..106T}, which are suggested from Fermi studies on diffuse $\gamma$-ray emission in the 2nd Galactic 
  quadrant \citep{2009arXiv0907.0312T}. 
  The accelerated spectrum of primary CRs at source region is assumbed to be a broken power law function:

\begin{equation}
   q(p)  =  q_0\times \left \{ 
     \begin{array}{ll}
          (p/p_{br})^{-\nu _1} & if (p<p_{br}), \\
          (p/p_{br})^{-\nu _2}\cdot f(\hat p)& if (p\geq p_{br})
     \end{array} 
   \right.
\label{CRSpetrum}
\end{equation}
  where p is the rigidity, $q_0$ is the normalization factor for all nuclei, relative abundance
  of each nuclei follows the default value in GALPROP package.
  $p_{br}$ is the broken energy and $\nu_1,\nu_2$ is the spectrum index
  before and after the broken energy $p_{br}$.
  $f(\hat p)$ is used to describe the high energy cut-off. For the primary electron,
  a soft spectrum index with the value of -3.5 at the rigidity 2 TV is adopted in order to agree with
  HESS observation \citep{2008PhRvL.101z1104A}. For the nuclei, to reproduce the knee structure of CRs,
  $f(\hat p)$ can be described as following formula 
  based on H\"{o}randel model \citep{2004APh....21..241H}:
\begin{equation}
f(p_{knee}) = \left[
   1+\left(\frac{p}{\hat p}\right)^{\epsilon_c}
   \right ]^{\frac{-\Delta\gamma}{\epsilon_c}}
\label{CRKnee}
\end{equation}
where $\Delta \gamma$ and $\epsilon_c$ characterize the change in
the spectrum at the break rigidity $\hat{p}$.
Detailed information of the parameters is listed in Table \ref{PrimaryCRs}.
\begin{table}[h]
\begin{center}
\caption{ The inject spectrum of primary CRs}
\begin{tabular}{ccc}\hline \hline
 parameters   & Nuclei & Electron  \\
 log($q_0$)   & -8.31 & -9.367     \\
 $\nu_1$      &  1.92  & 1.5       \\
 $\nu_2$      &  2.31  & 2.7       \\
 $p_{br}(GV)$ &  9     & 5.7       \\ \hline
 $\hat{p}(PV)$    &  3.68  &           \\
 $\Delta \gamma$ & 0.44 &          \\
 $\epsilon_c$ &  1.84   &          \\ \hline \hline
\end{tabular}
\label{PrimaryCRs}
\end{center}
\end{table}

  The ISM account for about 10$\%$-15$\%$ 
  of the total mass in the Galactic disk and its chemical composition is dominated by hydrogen and
  helium. The helium fraction is 11$\%$ by number density.
  The hydrogen gas $n_H$ includes three main components in
  molecular (H$_2$), atomic (H$_I$) and ionized (H$_{II}$ states. 
  The default gas distribution in GALPROP is adopted in our calculation, which is based on the survey results and 
  related modeling \citep{1988ApJ...324..248B,1976ApJ...208..346G,1991Natur.354..121C}.

\begin{figure}[!htb]
\centering
\includegraphics[width=0.48\textwidth]{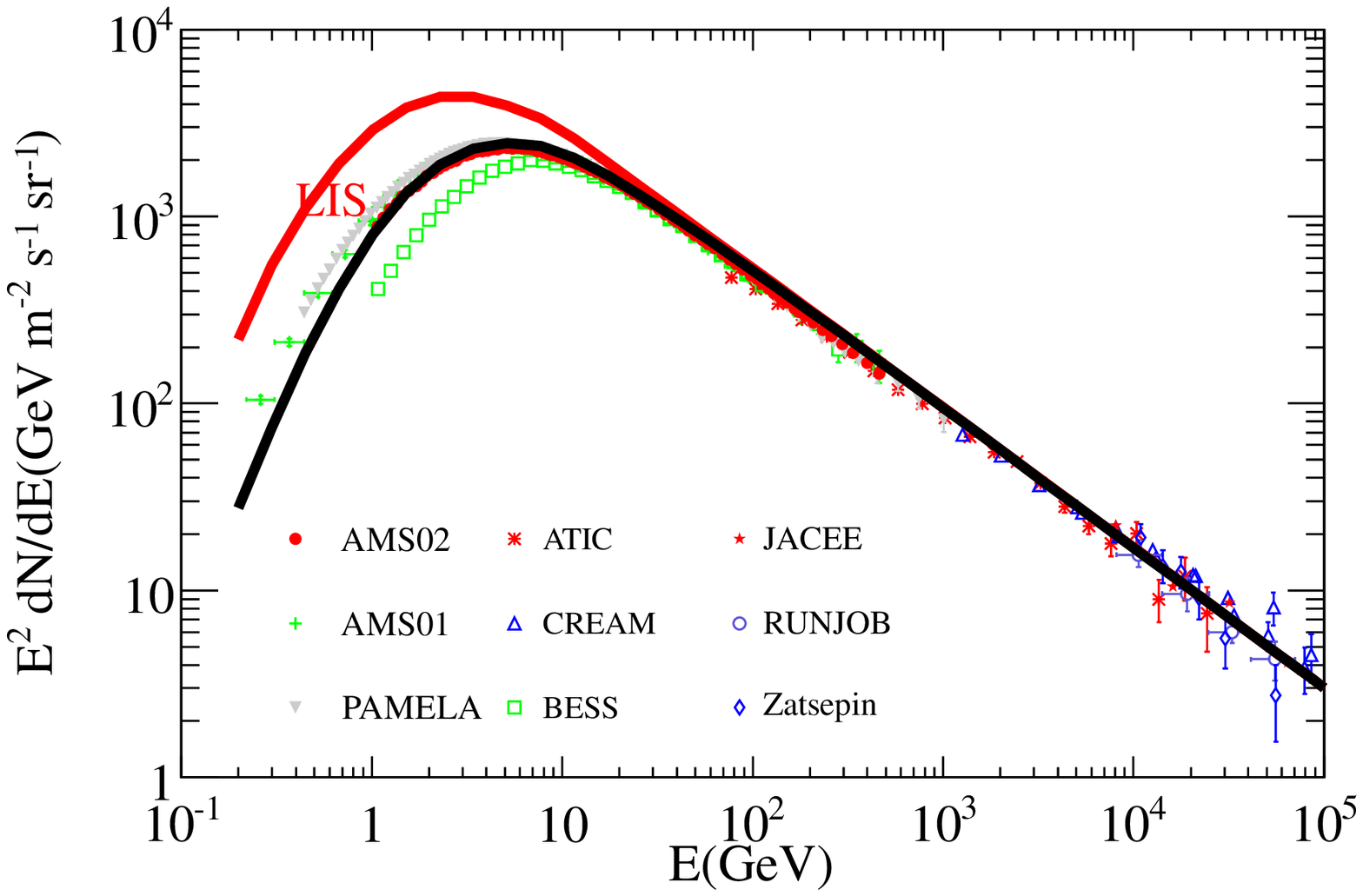}
\caption{The proton spectrum in the convention model. The experiment data of proton come from:AMS02\citep{2014arXiv1402.0467C},
         ATIC\citep{2006astro.ph.12377P}, PAMELA\citep{2011Sci...332...69A},AMS01\citep{2000PhLB..472..215A},
         CREAM\citep{2010ApJ...714L..89A},
         BESS\citep{2000ApJ...545.1135S},JACEE\citep{1998ApJ...502..278A},
         RUNJOB\citep{2001APh....16...13A},Zatsepin\citep{2006JPhCS..47...41H}}
\label{fig:PropProton}
\end{figure}

  By using publicly available numerical code GALPROP and taking the main parameters described above, 
  the directly measured proton spectrum up to $\sim$100 TeV can
  be successfully reproduced as shown in Fig. \ref{fig:PropProton}
  The solar modulation potential is assumed to 550 MV. Simultaneously, the calculation can provide all spectra of observable
  secondary particles for comparison with the experimental results. We will refer those results from correnponding calculation 
  as convention model results hereafter. 

  Convention model calculation can describe well the high latitude diffuse $\gamma$-ray 
  spectra but not enough for the spectra in the galactic plane \citep{2012ApJ...750....3A}.
  Distributed along the galactic disk, the fresh CRs has been discussed and expected to resolve 
  the deficit problem \citep{2009ApJ...695..883B,2014arXiv1412.1690N}.
  If CRs do have instant acceleration and injection, the propagation of 
  CRs does have an uniform diffusion parameters everywhere(inside and outside the galactic plane) and the CR sources do have
  continuous spatial and temporal distribution as
  assumed in GALPROP, then the solution of steady state can describe 
  the integrated CR distribution, including fresh and old CRs.
  However, the proceses of both CR acceleration and injection take time, and CR should diffuse
  slower in the galactic plane than in the halo. The CRs should spend longer time in galactic disk and have 
  effectively more interaction with ISM than what
  is calculated in GALPROP. Instead of using non-uniformly distributed diffusion coefficiency in the calculation,
  we simply add a fresh CR component in the source region to take care of the generation of additional secondary particles.
  One should be reminded that the fresh CR component does not take part in the propagation as the primary or secondary particles. 
  In this work, we assume that the fresh CRs have the same spatial distribution as that of the SNRs. Furthermore, 
  the spectrum of the fresh CRs should be close to the injected spectrum of primary CRs. As a matter of fact,
  we simply employ the injected spectrum of primary CRs as in Formula \ref{CRSpetrum}, but with a rescaled 
  normalization factor, which is determined by the detained time.  
  For example, if the fresh CRs take twice the time to escape from galactic disk than that of 
  a normal diffusion supposed by GALPROP calculation, the interaction 
  between the fresh CRs and the ISM would be doubled and the rescaled normalization factor would be one.
  In this work, the rescaled normalization factor is a free parameter and determined 
  by best accordance between the model calculation and the observation.

\section{Results on Secondary Particle Spectra}

  Base on above discussion, the secondary particles are produced from two components:
  steady state CRs in convention model and the fresh CRs. 
  The secondary spectra from the contribution of convention model 
  can be directly obtained from the GALPROP package calculation. 
  The secondary particle production from the fresh CRs is calculated 
  under the GALPROP frame by switching off the propagation of fresh CRs to 
  keep the fresh CR spectrum unchanged and limit the related interactions only in the source region. 
  After the production, the secondary particles
  follow the norminal propagation as those from conventional model calculation.
  With which, the spectra of additional secondary particles can be obtained.
  Given a rescaled normalization factor, the summed spectra of secondary particles can be compared with
  the observation. By adjusting the rescaled normalization factor, we find that a value of 0.4 offer the
  best agreement.

\subsection{Diffuse $\gamma$-ray emission}

  The diffuse $\gamma$-rays in the galactic plane are produced through
  three major processes: decay of $\pi^0$ generated by $pp$-collision, IC scattering off the ISRF and bremsstrahlung
  by electrons. In the case of IC calculation,
  the widely used ISRF model is adopted \citep{2000ApJ...537..763S,2005ICRC....4...77P}.

\begin{figure*}[!htb]
\centering
\includegraphics[width=0.9\textwidth]{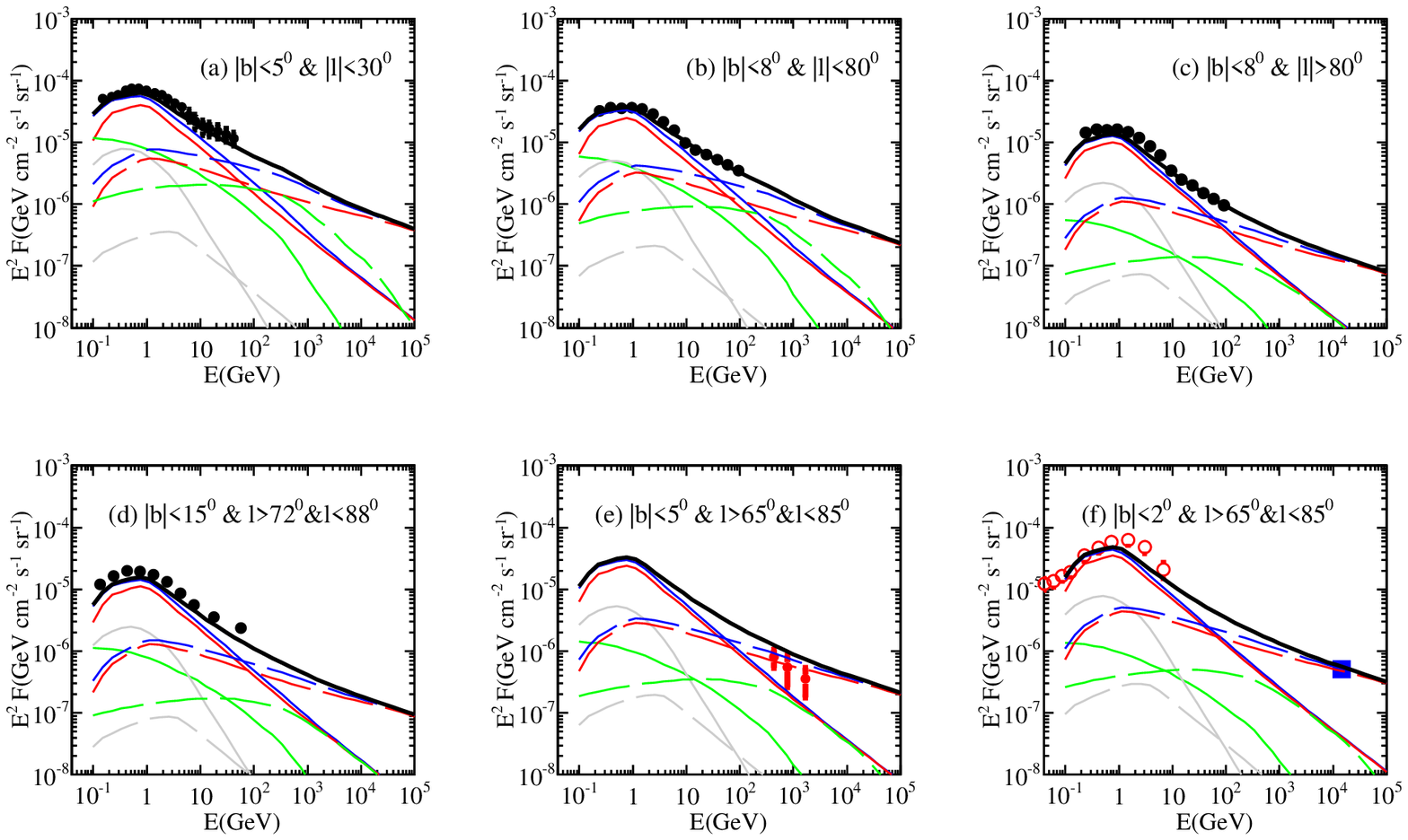}
\caption{The calculated $\gamma$-ray spectrum: 
         The red color shows the
         $\pi^0$ decay, green color shows the IC scattering, gray color shows the bremsstrahlung process and
         the black color shows the total one.
         The data at GeV energy range with black circle is from Fermi-LAT \citep{2014PhRvD..89f3515M,
         2010PhRvL.104j1101A,2012ApJ...750....3A,2012A&A...538A..71A} and with red open circle is from EGRET \citep{1997ApJ...481..205H}.
         The data at TeV energy range with red color points is from ARGO experiment \citep{ARGOSciascio};
         The data at 15 TeV energy with blue quadrangle points is from MILAGRO experiment \citep{2008ApJ...688.1078A}}.
\label{fig:gamma}
\end{figure*}

  According to the available spectra of high energy diffuse $\gamma$-ray on the galactic disk, 
  four regions are studied in this work:
  (a) inner most galactic plane($|b|<5^{\circ}$ \& $|l|<30^{\circ}$),
  (b) inner galactic plane ($|b|<8^{\circ}$ \& $|l|<80^{\circ}$), 
  (c) outer galactic plane($|b|<8^{\circ}$ \& $|l|>80^{\circ}$)
  and (d) CYGNUS region. 
  Fig. \ref{fig:gamma} (a)-(c) show the comparisons of the diffuse $\gamma$-rays from region (a)-(c).
  All of our convention model calculations agree well with those from Fermi-LAT collobration \citep{2012ApJ...750....3A}.
  The general conclusion is that model calculation can't describe the observation with the energy above 
  a few GeV. After adding the contribution from fresh CRs, also dominated by $\pi^0$ decay, 
  the agreements between model calculations and observations are much better improved from 1 to 100 GeV.
  The hard spectrum is expected to continue to very high 
  energy due to the hard spectrum of the fresh CR contribution, which can 
  be tested by diffuse $\gamma$-ray obsevation at higher energy.
  Fortunately, multi-TeV observation has been performed by ground-based EAS experiments in CYGNUS region.
  Fig. \ref{fig:gamma} (d1) show the spectrum observed by Fermi-LAT in a wide CYGNUS region together with model calculations.
  Though fresh CR helps to explain the observation above 10GeV, the overall theoretical spectrum underestimates the Fermi-LAT one.
  This is probably because that CYGNUS region is a star formation region which contains many accelerators, old and young.
  A full understanding should take into account all contributions, including these from point sources and extended sources.
  Another possibility is that the ISM in this region is not properly modeled in the calculation, if we increase the amount
  of gas by 25$\%$, the calculation can have a perfect agreement with the observation. ARGO-YBJ performed TeV observation
  in a slightly different area in CYGNUS region as shown in Fig. \ref{fig:gamma} (d2). The calculated flux agrees with
  the observation within experimental error. However, the measured flux has a large uncertainty and precision test 
  is clearly foreseen with new observation from Tibet-AS$\gamma$ \citep{2009APh....32..177S}, HAWC \citep{2013APh....50...26A}
  and future experiments such as LHASSO \citep{2010ChPhC..34..249C}, HiSCORE \citep{2012AIPC.1505..821T}.
  Fig. \ref{fig:gamma} shows the spectra meaurement by EGRET and MILAGRO from a narrow band of CYGNUS region
  along galactic plane. Model calculation shows very good agreement with the observation from sub-GeV to tens of TeV energy.
  The fresh CRs contribute almost all the $\gamma$-ray emission in multi-TeV energy. 
  More accurate observations of diffuse $\gamma$-rays above multi-TeV energy will be crucial in testing the fresh CR hypothesis.


\subsection{Diffuse neutrino emission}

  The charged pion decay will produce neutrinos accompaning with the
  $\gamma$-rays. 
  Different from $\gamma$-ray observation, neutrinos interact very weakly with matter and
  very large volume of target material is required to detect the neutrino events.
  
\begin{figure*}[!htb]
\centering
\includegraphics[width=0.48\textwidth]{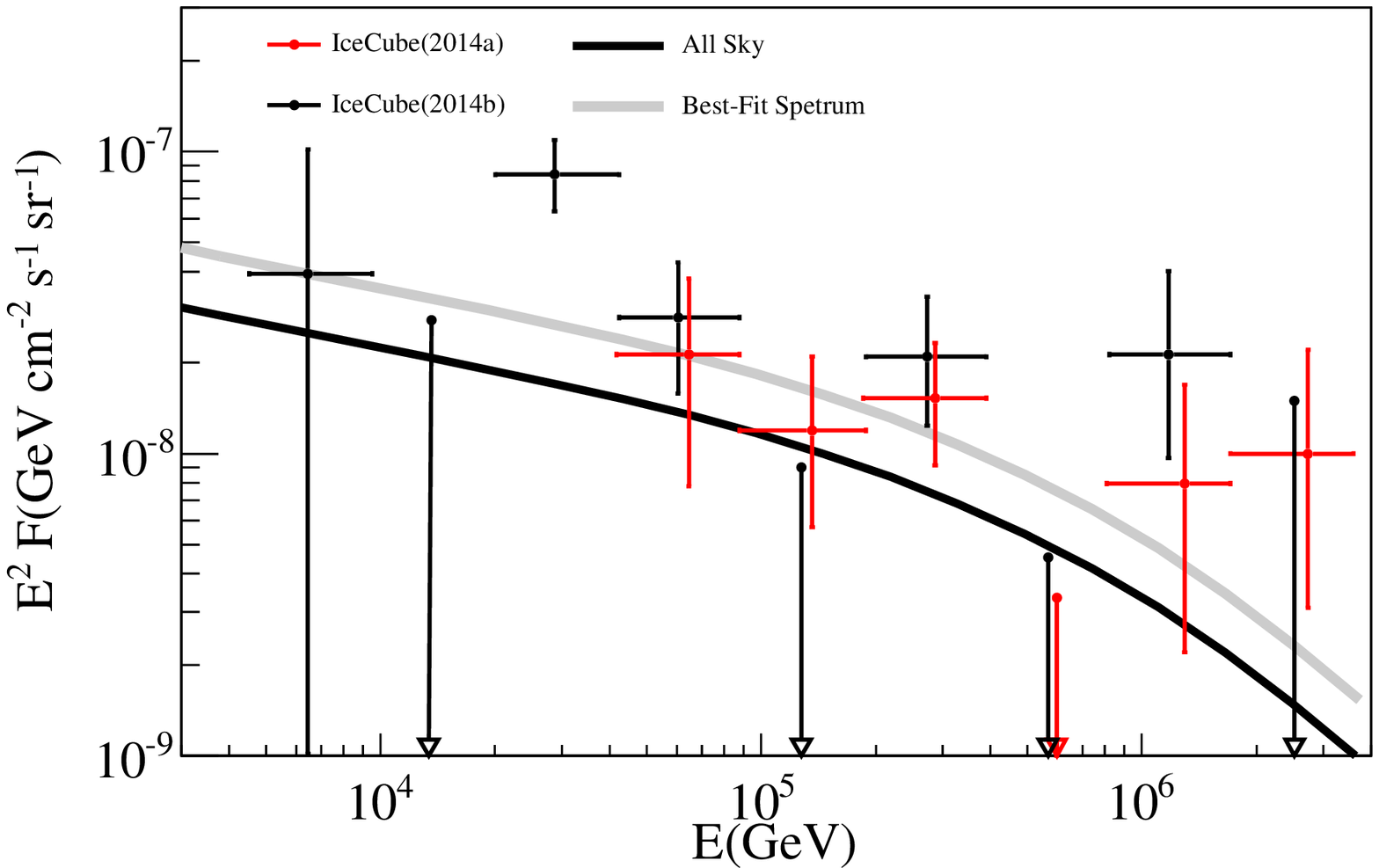}
\includegraphics[width=0.48\textwidth]{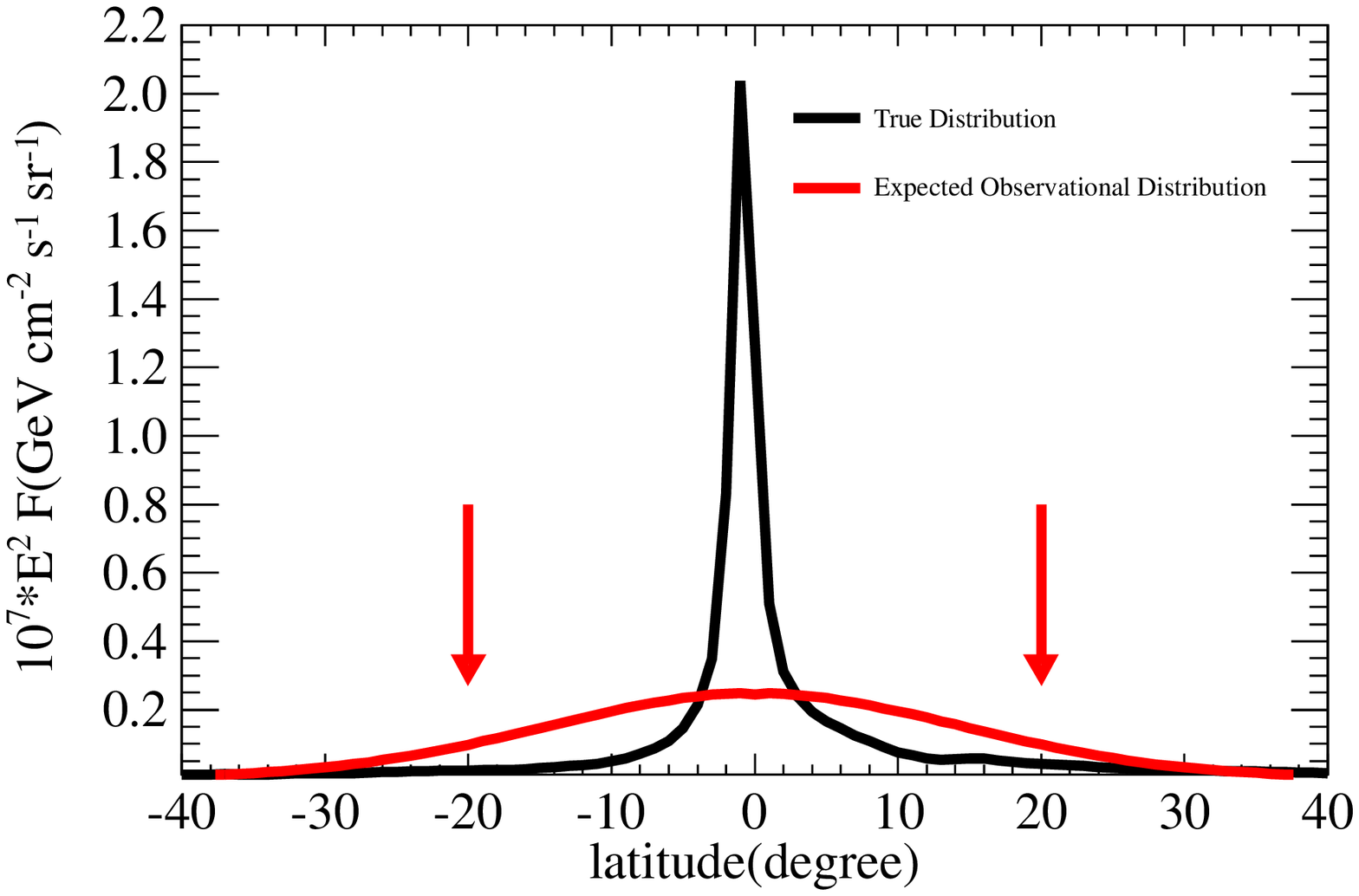}
\caption{The calculated diffuse neutrino spectrum(left panel) from collision of the
         CRs with the ISM. The data is astrophysical neutrino 
	 observation \citep{2014PhRvL.113j1101A,2014arXiv1410.1749A}. The righ panel is the integrated 
         neutrino flux for E$>$30 TeV along l=0. The black color is the true direction in our model and
         the red color is the recontructed direction after considering the angle resolution of 15$\%$.}
\label{fig:neutrino}
\end{figure*}

  With a cubic kilometer ice telescope below the surface of the South Pole, 
  the IceCube collaboration reported a detection of 37 neutrino candidate events from
  30 TeV to 2 PeV with a backgrounds of $8.4\pm4.2$ from CR muon events 
  and $6.6^{+5.9}_{-1.6}$ from atmospheric neutrinos during a livetime of 988 days \citep{2014PhRvL.113j1101A}.
  By including TeV measurement, the neutrino spectrum is obtained from $\sim$TeV to $\sim$PeV 
  energy \citep{2014arXiv1410.1749A}. The left panel of Fig. \ref{fig:neutrino} shows the 
  IceCube data points together with model calculation.
  It can be seen that the theoretical calculation of allsky flux in black solid line is lower than the experiment observation.
  The best-fitted result, shown in gray solid color, indicate that the fresh CRs contribute $\sim$60$\%$ of 
  IceCube observation.
  It should be noted that the galactic neutrino flux mainly come from galactic plane according
  to our model calculation as shown in the right panel of Fig. \ref{fig:neutrino} in black solid line.
  This conclusion apparently controdict with that of IceCube collaboration , which claimed an isotropic distribution 
  based on current limited number of neutrino events. To decribe the isotropic neutrino distribution, extra-galactic
  contribution is necessary. In fact, the neutrino flux from extra-galaxy is constrained by the measurements of 
  extra-galactic $\gamma$-ray background (EGB) \citep{2010PhRvL.104j1101A,2014arXiv1410.3696T} and expected to be at 
  the level of 10$\%$-50$\%$ contribution from star forming galaxies \citep{2014arXiv1412.1690N}, which is consistent with
  the part uncovered by the fresh CRs.

  Nevertheless, we can compare the number of neutrino calculation and observation 
  as listed in Table \ref{NeutrinoCalc}. $N_{theo}$ is calculated neutrino numbers in our model after considering
  the $15^\circ$ angular resolution of IceCube. We first smear the neutrino distribution by $15^{\circ}$ to obtain
  the expected observational distribution as shown by the red solid line in the right panel of Fig. \ref{fig:neutrino}.
  Then, we integrate the neutrino number distribution with latitude coodinate from $-20^{\circ}$ to $20^{\circ}$ 
  as indicated by the arrows 
  in the right panel of Fig. \ref{fig:neutrino} and get the $N_{theo}$.
  $N_{bkg}$ is the background numbers, estimated by multiplying the number of event fraction in choosen region
  with the total background numbers(8.4+6.6) assuming an isotropic background distribution.
  $N_{obs}$ is the neutrino numbers observed by IceCube experiments in the correspondent region.
  It is interesting to note that the summation of $N_{theo}$ and $N_{bkg}$ agree with $N_{obs}$ very well
  for four regions in the galactic plane. The four regions have a same defination as that were used for
  comparison of diffuse $\gamma$-ray spectra in previous section, except the latitude interval is fixed
  to $|b|<20^\circ$ to take into account the IceCube angular resolution.

\begin{table}[h]
\begin{center}
\caption{The comparison between theoritical calculation and the observation of IceCube \citep{2014PhRvL.113j1101A}}
\begin{tabular}{|c|c|c|c|c|}\hline \hline
 Region  & $N_{theo}$ &$N_{bkg}$& $N_{obs}$&Event Number \\ \hline
 Inner most  &3.0 &0.9  & 5 &$\#$2,$\#$14,$\#$22\\ 
 Galactic Plane        &   &  && $\#$24,$\#$25,  \\ \hline
 Inner Galactic                &  &   &  & $\#$2,$\#$4,$\#$14,$\#$22  \\
 Plane  &6.4& 2.4  &10& $\#$24,$\#$25,$\#29$,   \\ 
        &          &   && $\#33$, $\#34$, $\#35$   \\ \hline
 Outer Galactic  &3.4 &3 &5&$\#13$,$\#3$,$\#6$, \\
  Plane                 &   &  &&  $\#27$,$\#5$  \\ \hline 
Cygnus Region  &0.6  & 0.4 &2& $\#$29,$\#$34   \\ \hline \hline
\end{tabular}
\label{NeutrinoCalc}
\end{center}
\end{table}

  
\subsection{The ratio of $\overline p/p$ and $B/C$}

  There is a great astrophysical interest in CR antiprotons and Borons.
  It is believed that most of antiprotons and Borons are originated from collisions
  of CR with ISM.
  Therefore the information of CR propagation can be extracted from the comparison between 
  the spectra of secondary particles with those of primary CRs.
  Base on the above discussion, the fresh CRs can explain the diffuse $\gamma$-ray excess,
  and part of the IceCube neutrinos. It is no doubt that the secondary particle
  antiproton and Boron should exhibit excesses for energy above tens of GeV correspondingly.

\begin{figure*}[!htb]
\centering
\includegraphics[width=0.47\textwidth]{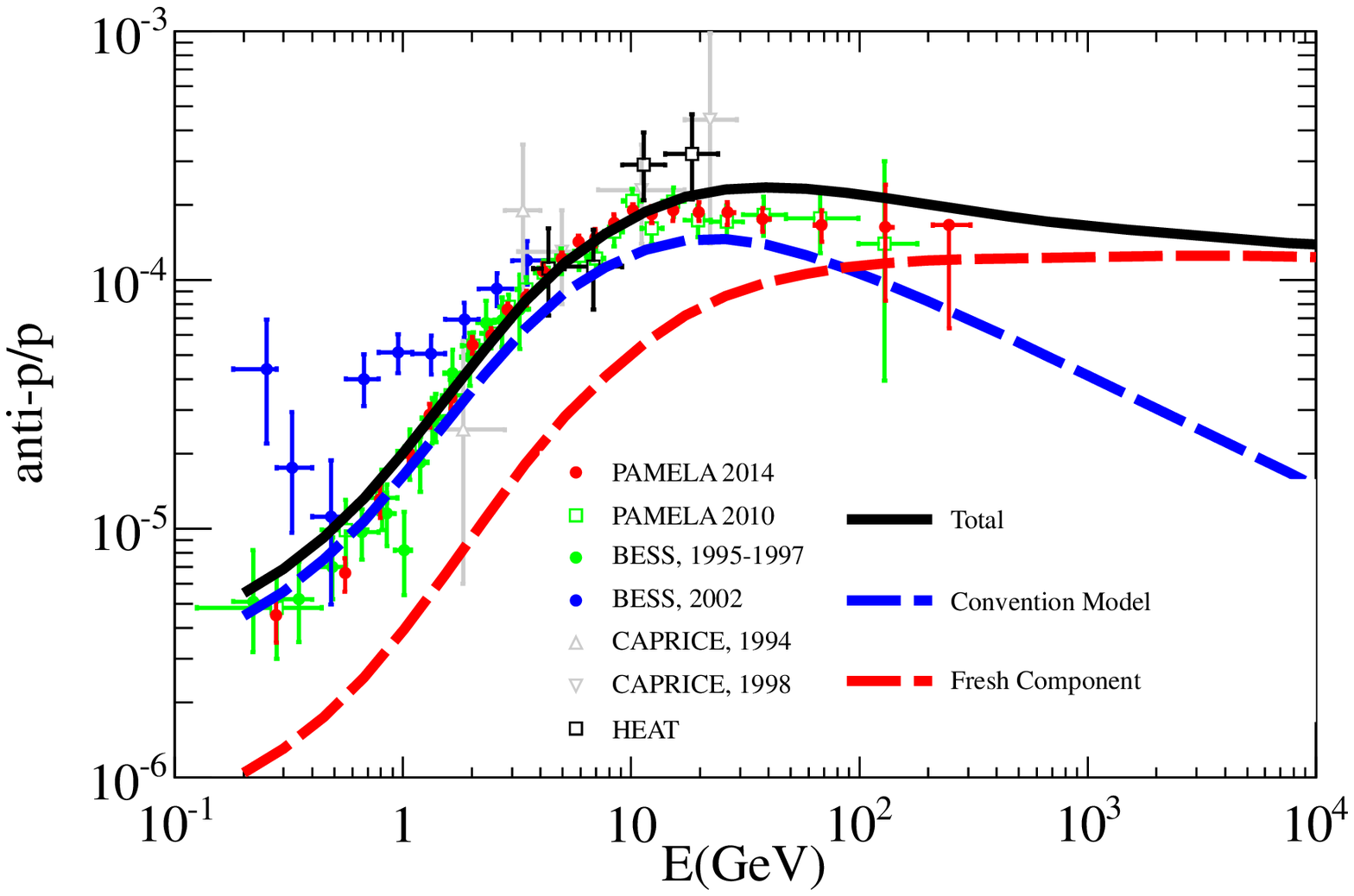}
\includegraphics[width=0.47\textwidth]{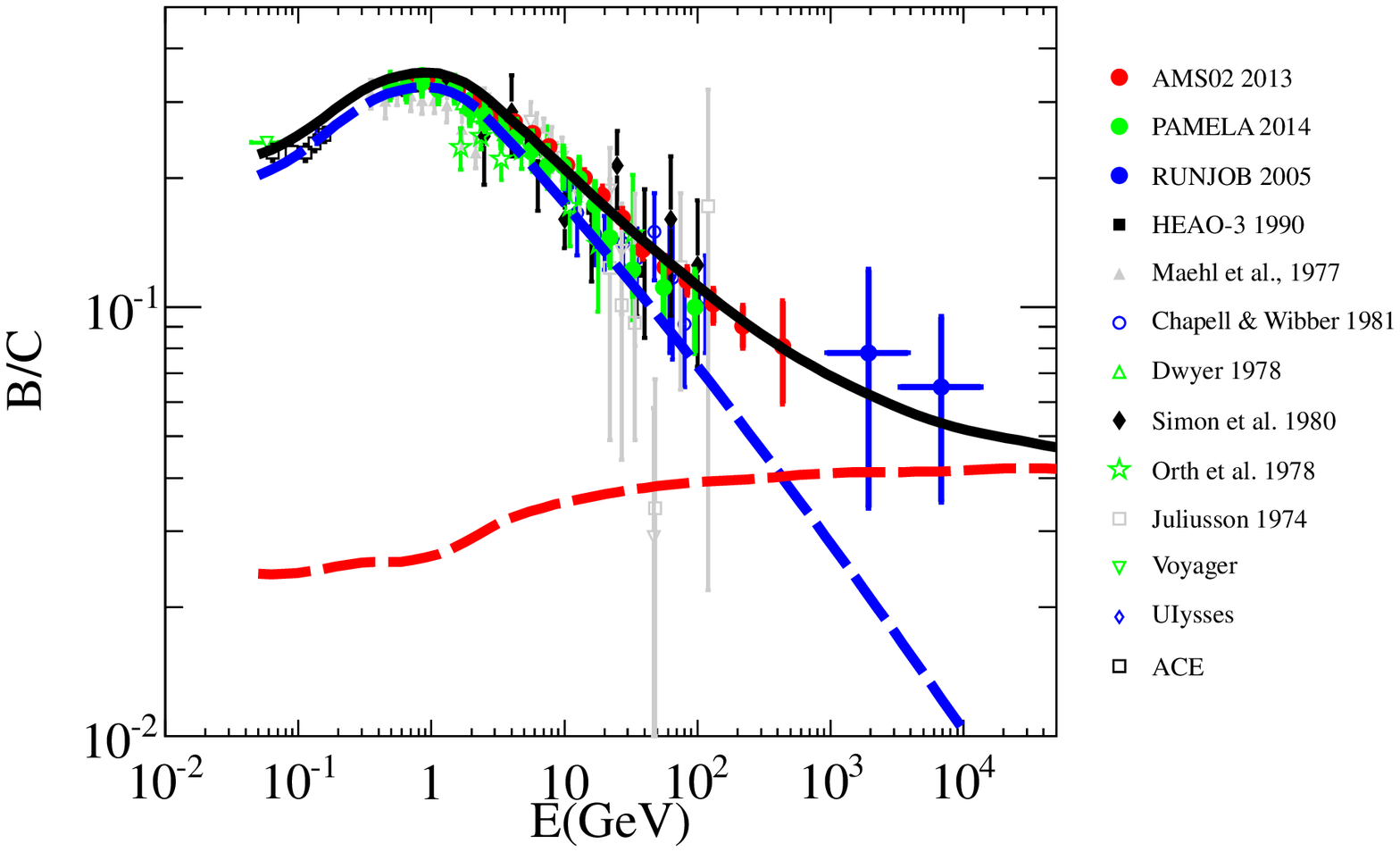}
\caption{The calculated $\overline p/p$ (left panel) and B/C (right panel). The $\overline p/p$ data
         from: PAMELA 2014 \citep{2014PhysicsReport},
         PAMELA 2010 \citep{2010PhRvL.105l1101A}, BESS 1995-1997 \citep{2000PhRvL..84.1078O}, BESS 1999 \citep{2002PhRvL..88e1101A},
         CAPRICE 1994 \citep{1997ApJ...487..415B}, CAPRICE 1998 \citep{2001ApJ...561..787B}, HEAT \citep{2001PhRvL..87A1101B}.
         The B/C data from: AMS02 \citep{AMS02BC}, PAMELA \citep{2014ApJ...791...93A}, RUNJOB \citep{2005ApJ...628L..41D},
         Juliusson \citep{1974ApJ...191..331J}, Dwyer \citep{1978ApJ...224..691D},
         Orth \citep{1978ApJ...226.1147O}, Simon \citep{1980ApJ...239..911S},
         HEAO-3 \citep{1990A&A...233...96E}, Maehl \citep{1977Ap&SS..47..163M}, Voyager \citep{1999ICRC....3...41L},
         Ulysses \citep{1996A&A...316..555D}, ACE \citep{2000AIPC..528..421D} and for other references see \citep{1998ApJ...505..266S}.}
\label{fig:BC}
\end{figure*}

  The left panel of Fig. \ref{fig:BC} shows the calculated $\overline p/p$ and the right panel shows the $B/C$.
  The blue and red dashed line presents the calculated results from conventional and fresh
  CR components respectively. The black solid line gives the total contribution. 
  Considering the contribution of fresh CRs, the overall calculation on the ratio of $\overline p/p$
  is a little higher than PAMELA observation \citep{2010PhRvL.105l1101A,2014PhysicsReport}. 
  However, the uncertainty of $\overline p$
  production cross section is $\sim 25\%$ over the energy 
  range 0.1 - 100 GeV \citep{2001ApJ...563..172D,2009PhRvL.102g1301D}, which should lead to 
  a same level of uncertainty in the ratio calculation.
  Taking all of those factors into account, the model calculation is consistent 
  with the observation within the errors. 
  On the other hand,
  the ratio of $B/C$ is quite consistent with AMS02 observation after considering the fresh CR contribution.
  Owing to its hard spectrum, the fresh CRs induced secondary $\overline p$ and Boron inherit a similar
  hard spectrum, which make the ratio of $\overline/p, B/C$ considerably flater than 
  that from conventional model. Such a tendency is not obvious in current observation.
  In the future, high statistic and TeV energy observation can offer a crucial and definitive identification of this model.


\subsection{Positron and Electron Excess}
 
  The charged pion decay will 
  produce $e^+e^-$ accompanying with the neutrino. In this section, 
  the contribution of secondary particle
  $e^+e^-$ from the fresh CRs interaction with ISM will be discussed. 

\begin{figure*}[!htb]
\centering
\includegraphics[width=0.48\textwidth]{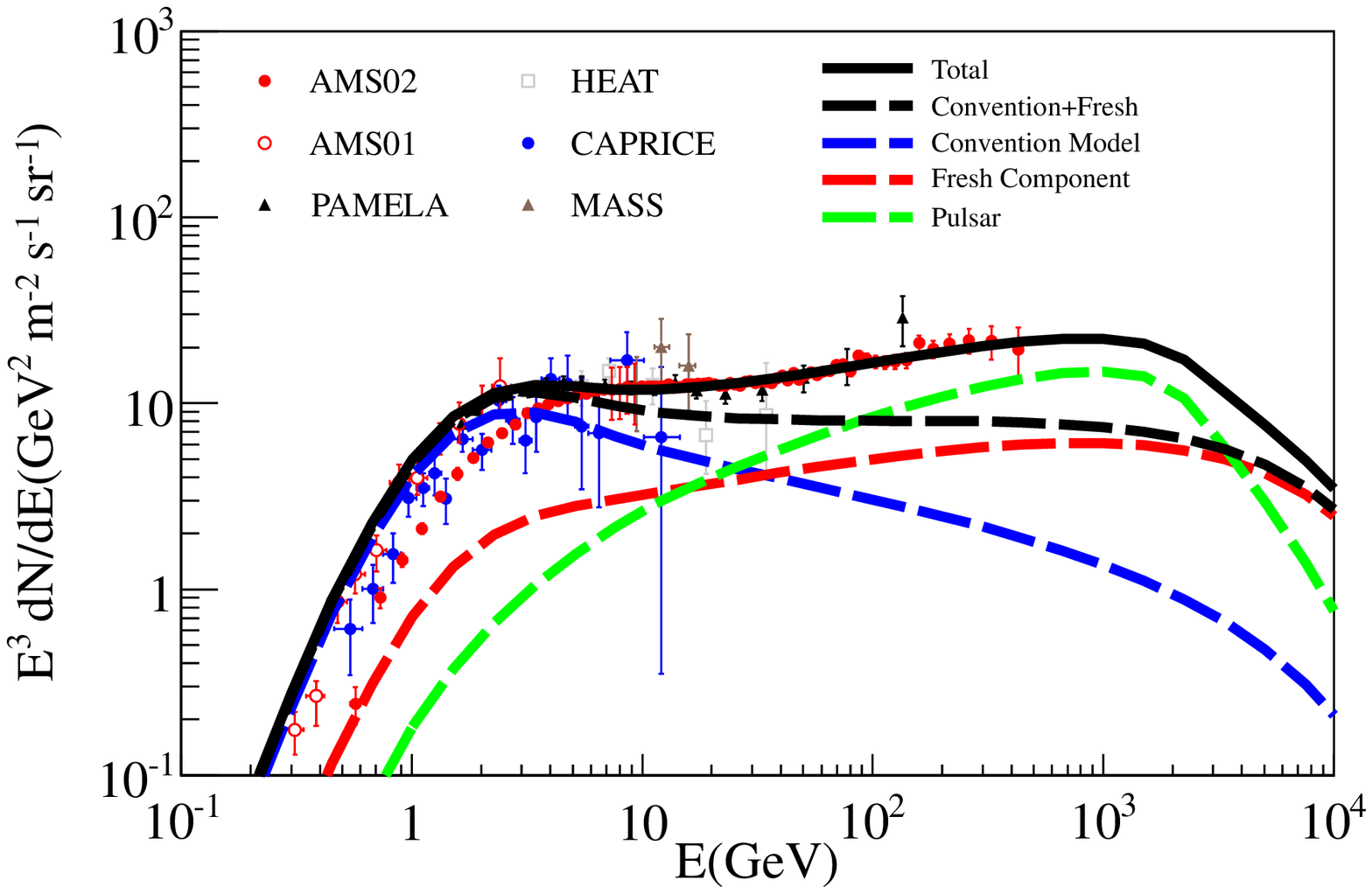}
\includegraphics[width=0.48\textwidth]{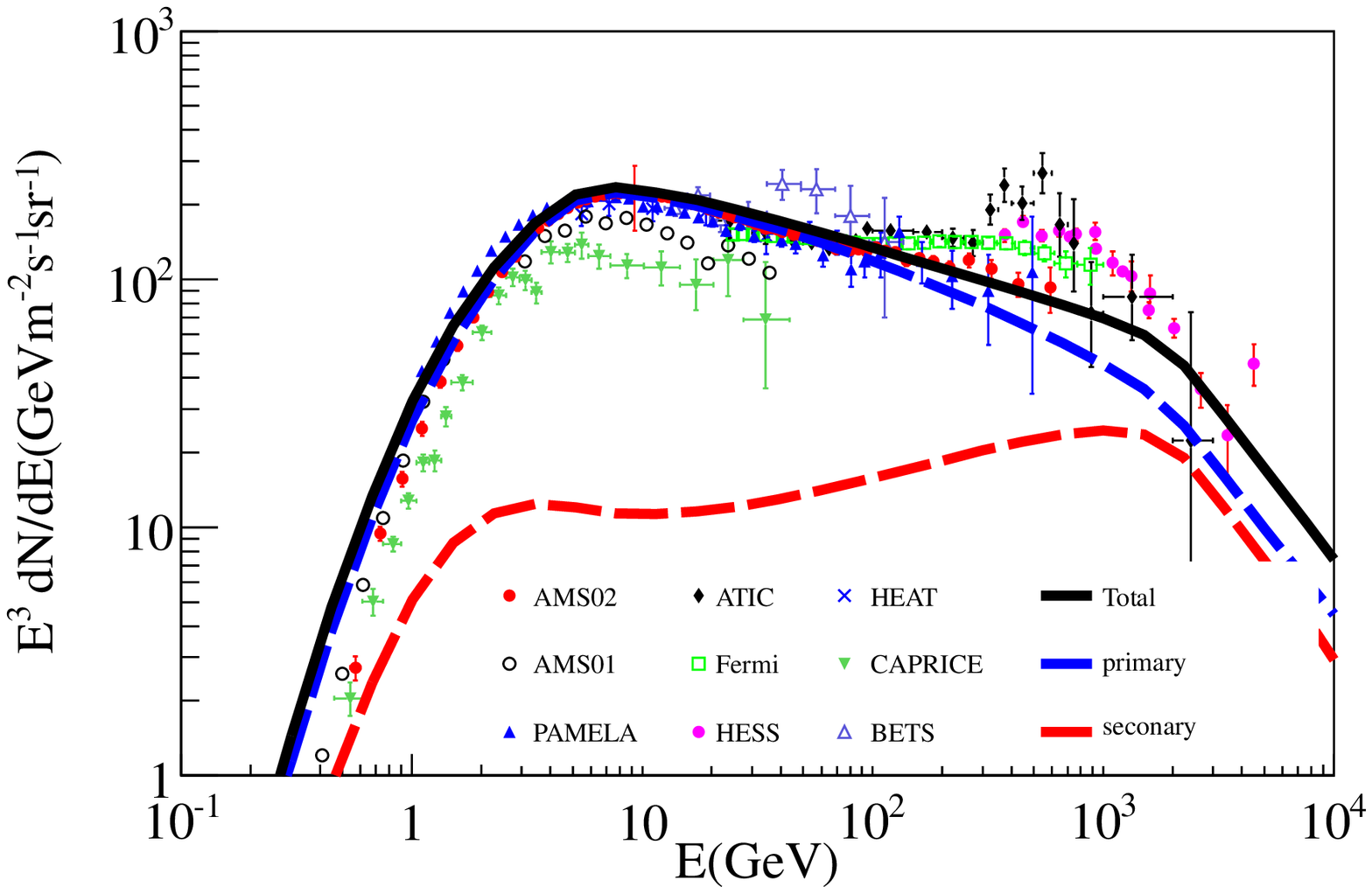}
\caption{The calculated positron (left panel) and electron (right panel) spetrum. The experiment datas are
         adopted from AMS02 \citep{2014PhRvL.113l1102A}, AMS01 \citep{2000PhLB..484...10A}, 
         PAMELA \citep{2009Natur.458..607A,2011PhRvL.106t1101A}, 
         HEAT \citep{1998ApJ...498..779B}, CAPRICE \citep{2000ApJ...532..653B},
         ATIC \citep{2008Natur.456..362C}, Fermi-LAT \citep{2010PhRvD..82i2004A}, HESS \citep{2008PhRvL.101z1104A} 
         and BETS \citep{2001ApJ...559..973T}.}
\label{fig:LeptonExcess}
\end{figure*}

  The left panel of Fig. \ref{fig:LeptonExcess} shows the positron spectra from two contributions.
  The blue and red dash line stand for the convention model and fresh CR calculation. 
  Because of the energy loss during the propagation of positron, the resulting spectrum of positron
  from fresh CRs become softer, which make the sum of the two contributions not enough to explain
  the positron excess as shown by the black dash line.
  The pulsar sources or exotic physical process is required to account for the discrepancy.
  According to the pulsar model \citep{2015APh....60....1Y}, positron spectrum from pulsar can be described 
  by formula \ref{CRSpetrum}.
  Combined the three parts of
  components, the total calculation agree with AMS02 observation well as shown in black solid line. 

  The right panel of Fig. \ref{fig:LeptonExcess} contains the spectrum of electron from observations
  and calculations. The blue dash line is for the primary electron from the acceleration sources and
  the red dash line is the summation of three contributions as for case of positron.
  In total, they agree with the spectra measured by PAMELA and AMS02 as shown in black solid line.

\begin{figure}[!htb]
\centering
\includegraphics[width=0.49\textwidth]{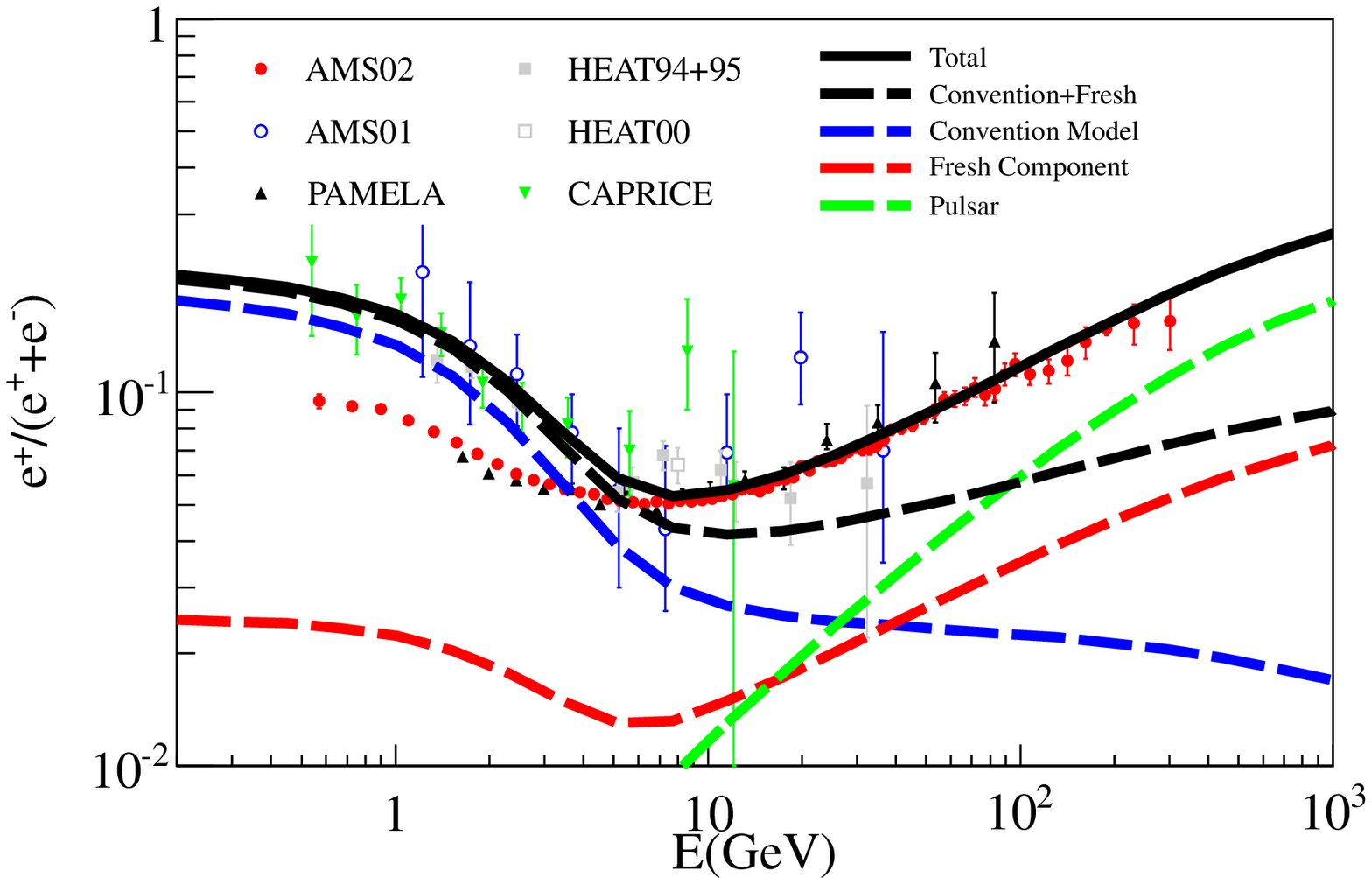}
\caption{The calculated positron fraction. The experiment datas are adopted from AMS02 \citep{2013PhRvL.110n1102A},
         AMS01 \citep{2007PhLB..646..145A}, PAMELA \citep{2009Natur.458..607A}, 
         HEAT94+95 \citep{1997ApJ...482L.191B}, HEAT00 \citep{2001ICRC....5.1687C} and CAPRICE \citep{2000ApJ...532..653B}.}
\label{fig:RatioLepton}
\end{figure}

  Because of the good agreement between the model calculation and observation for both positron and
  electron spectra, without surprise our model describe very well the ratio of positron to electron
  as shown in Fig. \ref{fig:RatioLepton}.

\section{Discussion and Conclusion}

  Origin, acceleration and propagation are the fundamental problems of CRs physics. In an era when high 
  precision and multi-messenger observation results keep flood in, when various excesses and new phenomena 
  continually pop up, the successful standard model of CRs is facing stringent test and refinement. 
  Primary spectra are the most important inputs for the stand model and have decisive effect to the spectra of 
  secondary particles. While primary spectra measured by PAMELA and CREAM hint that galactic CRs may 
  have a hard component at the source and can be responsible for the diffuse $\gamma$-ray and neutrino excesses. 
  More accurate observation made by AMS02 apparently disapproves 
  the rapid hardening of primary nucleus spectra around 200 GV rigidity.
  To recover the production of the hard spectra for diffuse $\gamma$-rays in 
  galactic plane, an alternative solution is that the fresh CRs might have stayed longer in galactic disk 
  than we previous have thought.

  Instead of incorporating a spatial dependent diffusion coefficient in the calculation, by simply adding a fresh 
  CR component to account for the additional secondary particles production, the diffuse $\gamma$-ray excess 
  from 10 GeV to muti-TeV energy can be successfully explained. According to which, we found that the fresh CRs
  should spend about 40$\%$ longer time in galactic plane than that assumed by standard model. Same process can generate 
  additional neutrino and explain about half of the Ice Cube neutrino flux. However, just like diffuse $\gamma$-rays, 
  these diffuse neutrinos lie on the galactic plane and have a high level of anisotropic distribution. To fully 
  understand the isotropic Ice Cube neutrino, extra-galactic contribution is inevitable. Though the fresh CRs 
  can generate right amount of electron and positron, but they are not able to explain the total electron and positron 
  excess simply because of the energy loss on the journey of propagation. If electron and positron can undergo a fast 
  diffusion with less energy loss, we expect that fresh CRs can fully account for the excesses. In current 
  scenario, electron and positron contribution from astrophysics sources, such as pulsar is necessary. A very 
  important difference between pulsar and fresh CRs is that the latter model predicts additional production 
  for all secondary particles, including anti-proton and Boron. According to our calculation, the ratio of $\overline p/p$ 
  and $B/C$ will become flatter for energy above tens of GeV. High precision observation of 
  flat ratios from AMS02 will be the smoking gun to test the fresh CR model. From theoretical point of view, 
  a spatial dependent diffusion coefficient may lead to an observable hard component in primary CRs, further 
  theoretical study is needed probably under the frame of DRAGON \citep{2008JPCM...20U2205G}.

\section*{Acknowledgements}
  We thank XiaoJun Bi and Qiang Yuan for helpful discussion.
  This work is supported by the Ministry of Science and Technology of
  China, Natural Sciences Foundation of China (11135010).

\bibliographystyle{apj}
\bibliography{hybrid}

\end{document}